\documentclass[twocolumn,showpacs,preprintnumbers,amsmath,amssymb]{revtex4}

\usepackage{graphicx}

\usepackage{dcolumn}
\usepackage{bm}

\newcommand{\be}{\begin{equation}}
\newcommand{\ee}{\end{equation}}
\newcommand{\ba}{\begin{align}}
\newcommand{\ea}{\end{align}}
\newcommand{\bea}{\begin{eqnarray}}
\newcommand{\eea}{\end{eqnarray}}
\newcommand{\rd}{\rm{d}}
\newcommand{\rg}{\rm{g}}

\makeatletter
\def\elsartstyle{%
    \def\normalsize{\@setfontsize\normalsize\@xiipt{14.5}}
    \def\small{\@setfontsize\small\@xipt{13.6}}
    \let\footnotesize=\small
    \def\large{\@setfontsize\large\@xivpt{18}}
    \def\Large{\@setfontsize\Large\@xviipt{22}}
    \skip\@mpfootins = 18\p@ \@plus 2\p@
    \normalsize
}
\@ifundefined{square}{}{}
\makeatother

\begin{document}

\title{The Effect of Redshift Distortions on the Integrated Sachs-Wolfe Signal}

\author{Ana\"is Rassat}

 \email{anais.rassat@cea.fr}


\affiliation{IRFU-SAp, Service d'Astrophysique, CEA-Saclay, F-91191 Gif sur Yvette Cedex, France.\\
}%

\date{\today}

\begin{abstract}
We show that  linear redshift distortions in the galaxy distribution can affect the ISW \emph{galaxy}-temperature signal, when the galaxy selection function is derived from a redshift survey. We find this effect adds power to the ISW signal at all redshifts and is larger at higher redshifts.  Omission of this effect leads to an overestimation of the dark energy density $\Omega_\Lambda$ as well as an underestimation of statistical errors.  We find a new expression for the ISW Limber equation which includes redshift distortions, though we find that Limber equations for the ISW calculation are ill-suited for tomographic calculations when the redshift bin width is small.  The inclusion of redshift distortions provides a new cosmological handle in the ISW spectrum, which can help constrain dark energy parameters, curvature and alternative cosmologies.  Code is available on request and will soon be added as a module to the iCosmo platform (http://www.icosmo.org).
\end{abstract}

\maketitle

\section{Introduction}\label{intro}
The integrated Sachs-Wolfe effect \citep[ISW]{Sachs:1967er} is a purely linear effect (in the non-linear regime, it is the weaker Rees-Sciama effect) which in a flat universe is an independent signature of dark energy. Furthermore, as it probes both the expansion history of the Universe and the growth of structure, it can be used to constrain alternative models of gravity, as well as the curvature of the Universe.  These features make it a `clean' and attractive probe, even though its cosmological constraining power is weak compared to that of other probes such as galaxy clustering and weak lensing. 
 
Its relative weakness means future surveys are never specifically optimised to measure the ISW effect.  However \cite{douspis:2008} showed that the optimal survey for measuring the ISW effect is similar to one designed to study BAOs and weak lensing, so that measurement of the ISW effect effectively comes \emph{for free} with future LSS surveys, such as DUNE \citep{refregier:2008}, JDEM \footnote{http://jdem.gsfc.nasa.gov/} or Euclid \footnote{http://sci.esa.int/science-e/www/area/index.cfm?fareaid=102}.

Recent detections of the ISW effect by cross-correlation of WMAP data with tracers of LSS \citep{Fosalba:2003ge,Boughn:2004zm,Afshordi:2003xu,Nolta:2003uy,Padmanabhan:2004fy,Cabre:2006qm,Gaztanaga:2004sk,Giannantonio:2006al,McEwen:2006md,Rassat:2007krl} show it is a promising cosmological probe, especially where a tomographic study is possible \cite{Giannantonio:2008}, as will be the case with future surveys.  However, some of the studies above \citep{Cabre:2006qm,Rassat:2007krl,Giannantonio:2008} return best fit values for the dark energy density between $\Omega_\Lambda = 0.80-0.85$, i.e., higher than that expected by today's concordance cosmology.  This possible discrepency is currently not explained, although it could simply be due to cosmic variance.

It was also originally thought that measuring the correlation between LSS and the CMB was a direct measure of the ISW effect \citep{Crittenden:1995ak}; \cite{Loverde:2007} showed that such correlations also include a cosmic magnification signal which may mimic or dampen the ISW effect, especially at high redshifts.  Rather than a hindrance to ISW measurements, such extra correlations are encouraging as they make the LSS-CMB correlation signal less featureless than previously thought, which means degeneracies between different cosmological parameters are more likely to be broken. It also provides a handle on dark energy at higher redshifts.

Another promising probe are linear redshift distortions \citep*{Fisher:1994,Heavens:1995,Guzzo:2008}, which are present in the galaxy density field. These arise from the galaxy peculiar velocity field, which itself correlates with the CMB's ISW signal \citep{Dore:2007}.  Though in principle the velocity-temperature signal to noise is larger than the galaxy-temperature signal to noise, large velocity maps are much harder to create than galaxy maps.  

In this paper we show that these redshift distortions can also contribute to the \emph{galaxy}-temperature cross-correlation signal.  In section \ref{origin} we review the origin of the ISW signal.  In section \ref{linear} we give the linear theory expression for the ISW cross-power, which we extend to include the effect of redshift distortions in section \ref{redshift}.  In section \ref{limber} we derive a new Limber equation for the ISW cross-power which includes the effect of redshift distortions.  In section \ref{ccl} we present our main conclusions.

\section{The Origin of the ISW Effect}\label{origin}
The gravitational potential of the LSS distorts space-time so that a photon travelling 
through it will be subject to a gravitational blueshift on entry of the potential and 
redshift on exit. If the potential does not vary during the photon travel time, then the net effect will be null: the photon will emerge unaffected 
by LSS. This is always the case on linear scales in an Einstein-de Sitter Universe.

If, however, these  potential wells vary with time, as they would in the 
presence of dark energy or curvature, the photon will emerge from the LSS gravitational 
field, either red- or blue-shifted depending on whether the potentials grow or decay respectively.

For photons travelling from the surface of last scattering, the varying gravitational potential of LSS will create secondary temperature anisotropies which will add 
power to the temperature-temperature (T-T) angular power spectrum $C_{TT}(\ell)$. The power added 
on large scales in the case of non-anisotropic stress is \citep{Sachs:1967er}: 
\be \left(\frac{\Delta T}{T}\right)_{ISW}= -2 \int_{\eta_L}^{\eta_0}\Phi'\left(\left(\eta_0-\eta\right){\bf \hat{n}},\eta\right)d\eta,\ee
where $T$ is the temperature of the CMB, $\eta$ the conformal time, defined by $d\eta=dt a(t)$ and $\eta_0$ and $\eta_L$ are the conformal times today and at the surface of last scattering respectively; 
${\bf \hat{n} }$ is the unit vector along the line of sight; $\Phi(x,\eta)$ is the gravitational potential at position $x$ and at conformal time $\eta$ and $\Phi' \equiv \frac{\partial \Phi}{\partial \eta}$.  The factor $2$ arises from assuming the Newtonian potentials $\psi$ and $\Phi$ are equal.

The ISW signal is weak compared to primary temperature anisotropies which makes it difficult to extract from the T-T power spectrum alone, though recent methods to reconstruct the actual ISW T-T signal exist \citep{iswmap:2008, granett:2008}.

Current detections of the ISW signal use a method proposed by \cite{Crittenden:1995ak} which measures the cross-correlation between LSS and the CMB to detect the ISW effect independently from the intrinsic CMB fluctuations. A significant decay in the gravitational potentials will produce large scale hot spots in the CMB. These gravitational potentials will also tend to host an overdensity of galaxies, so a positive correlation between the CMB and the galaxy distribution is expected. 

\section{Linear Theory Predictions}\label{linear}

The predicted cross-correlation signal of the ISW effect in spherical harmonic space is given by: 
\be C_{\rg T}(\ell)=4\pi\int \rd k \frac{\Delta^2(k)}{k} W_{\rg,
  \ell}(k)W_{T,\ell}(k)\label{linear:eq:cgt}\ee
where $\Delta^2(k) = \frac{4\pi}{(2\pi)^3}k^3P(k).$   Eq. \ref{linear:eq:cgt} is the exact equation in linear theory.
The temperature window function is give by: 
\be W_{T,\ell}(k)=A_k\int_0^{z_L} \rd r
j_\ell(kr)HD\left(\frac{\rm{dln}D}{\rm{dln}a}-1\right),\label{linear:eq:wt}\ee
where $A_k = \frac{-3\Omega_m^0H_0^2}{k^2c^3}$.  The terms $H$ and $D$ are the Hubble expansion and the growth function respectively, which depend implicity on redshift.  The last term in Eq. \ref{linear:eq:wt} shows that in a universe where $\frac{\rd \ln D(r)}{\rd \ln a} = 1$ at all redshifts (i.e. in an Einstein-de Sitter universe), there is no ISW signal expected.

In general the galaxy window function is taken to be: \be W^{\rm real}_{\rg,\ell}(k)=\int \rd r b(r)\varphi(r)j_\ell(kr)D.\label{linear:eq:wgreal}\ee This is the real (as opposed to redshift)
space window function. We assume a linear bias $b(r)$.  The term $\varphi(r)$ is the galaxy selection function, which represents the probability of finding a galaxy at a distance $r$ from the observer.

\begin{figure}
\includegraphics[width=8cm]{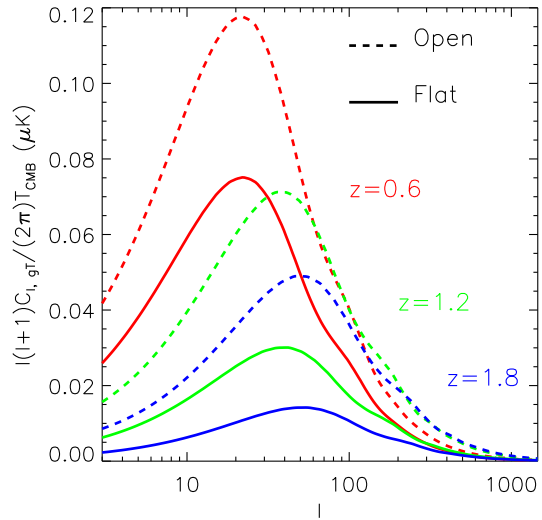}
\caption{The ISW signal for universes with open universes (dashed lines) compared to flat universes with the same matter content (solid lines).  The curvature density for the open universes is $\Omega_k = 0.10$, and the matter content is $\Omega_m  =0.25$.  The red, green and blue lines correspond to galaxy bins centered at $z=0.6, 1.2, 1.8$ resepectively, with $\Delta z=0.2$.  The amplitude of the ISW cross-correlation signal is always larger in open universes for all redshifts.}
\label{linear:fig:open}
\end{figure}

From Eq. \ref{linear:eq:wt}, we see that a positive ISW correlation (i.e., hot spots in the CMB corrspond to over-densities) is expected for universes where $\frac{\rd \ln D(r)}{\rd \ln a} > 1$, as is the case for open universes \citep{Kamionkowski:1994s,Kinkhabwala:1999k} as well as  universes containing dark energy.  In Figure \ref{linear:fig:open} we show that open universes present a larger ISW cross-correlation signal that flat universes with the same matter content.

\section{The Effect of Redshift Distortions on the ISW Signal}\label{redshift}
The galaxy window function given in Eq. \ref{linear:eq:wgreal} is valid in real space.   However, it is well known that a galaxy population whose radial distribution is estimated from redshifts will be subject to redshift distortions \citep*{Kaiser:1987,Fisher:1994,Heavens:1995}.  This will affect the radial galaxy power spectrum as well as the estimated selection function.  In the absence of redshift distortions, the redshift distribution of a galaxy is related to the population's selection function through: 
\be \varphi(r)= \frac{1}{\Omega r^2}\left(\frac{\rd N}{\rd z}\right)\frac{\rd z}{\rd r}.\ee
In the presence of linear redshift distortions, the inferred selection function will be distorted  such that \citep*{Fisher:1994}: 
\be \varphi(\tilde r) = \varphi(r) + \frac{d\varphi(r)}{\rd r}V(r),\label{redshift:eq:selectionz}\ee
where $V(r)$ is the radial peculiar velocity field of galaxies induced by their coherent motion on large scales, and Eq. \ref{redshift:eq:selectionz} assumes that the perturbations induced by the redshift distortions are small enough that a Taylor expansion of the selection function is valid.

Therefore, where the selection function is inferred from the measured redshift distribution of galaxies, the galaxy window function should include an extra term.  
Redshift distortions therefore affect the \emph{radial} galaxy power spectrum, as well as the \emph{projected} galaxy correlation  \citep{Padmanabhan:2006,Rassat:2008bao}. The extra term in the galaxy window function is given by \citep*[see][for a derivation]{Fisher:1994}: \be W_{\rg, \ell}(k)= W_{\rg,\ell}^{\rm {real}}(k) + \beta W_{\rg,\ell}^{\rm{z}}(k),\ee
where the distortion parameter $\beta$ is statistically related to the galaxy peculiar velocity field.  It modulates the amplitude of the redshift distortion effect and is defined by \citep{Peebles:1970Y}:
\be \beta = \frac{1}{b(z)}\frac{\rm{dln}D(z)}{\rm{dln}a} \simeq \Omega_m^\gamma(z).\ee
The window function of the redshift space contribution is given by:
\be W_{\rg,\ell}^{\rm{z}}(k)=\frac{1}{k}\int \rd r D\frac{d\varphi(r)}{dr}j'_\ell(kr),\label{redshift:eq:wgz}\ee where $j'_\ell(kr) \equiv \frac{\rd j_\ell(kr)}{\rd (kr)}$.  

 \cite{Padmanabhan:2006} also showed that Eq. \ref{redshift:eq:wgz} could be integrated by parts and
simplified, so that: 
\be W_{\rg, \ell}^{\rm{z}}(k)= \int \rd rD \varphi(r) \left[A_\ell
j_\ell(kr) - B_\ell j_{\ell-2}(kr) - D_\ell
j_{\ell+2}(kr)\right]\label{linear:eq:pad}, \ee
where $A_\ell= \frac{(2\ell^2+2\ell-1)}{(2\ell+3)(2\ell-1)}$, $B_\ell = \frac{\ell(\ell-1)}{(2\ell-1)(2\ell+1)}$ and $D_\ell = \frac{(\ell+1)(\ell+2)}{(2\ell+1)(2\ell+3)}$.

The redshift space window function is then: 
\be W_{g,\ell} (k) = W^{\rm real}_{g,\ell}(k)(1+\beta A_\ell)-\beta B_\ell W_{g,\ell-2}^{\rm{real}}(k)-\beta D_\ell W_{g,\ell+2}^{{\rm real}}(k)\label{redshift:eq:wgz:pad}\ee

This is the first time that the effect of redshift distortions on the galaxy window function has never been included in the ISW galaxy-temperature calculation.  The equation for the ISW effect, including the effect of redshift distortions is given by the same equation as before (i.e. Eq. \ref{linear:eq:cgt}), but where the galaxy window function is calculated using Eq. \ref{redshift:eq:wgz:pad}.  In this case, the ISW effect has a new dependence on the linear growth factor as well as on the linear bias, through the redshift distortion parameter $\beta$.

\begin{figure}
\includegraphics[width=8cm]{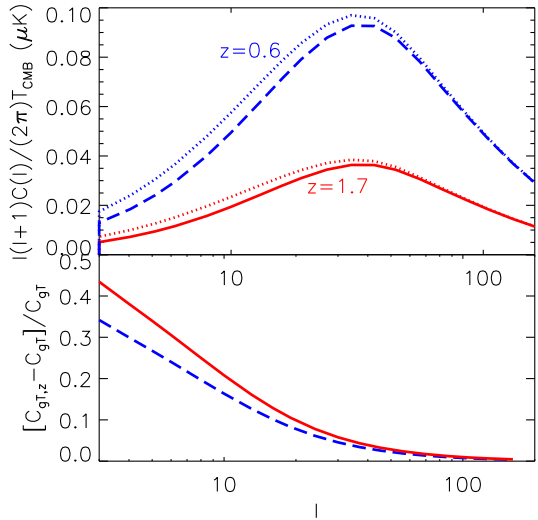}
\caption{The effect of redshift distortions on the ISW cross-correlation signal. \emph{Top panel}: the undistorted ISW cross-correlation signal for galaxy samples in bins centered at redshifts $z=0.6$ and $z=1.7$ (dashed and solid lines respectively), compared to the signal including the effect of redshift distortions (dotted lines). The effect of the redshift distortions is to add power to the ISW cross-correlation signal at all redshifts and on all scales. \emph{Bottom panel}: the relative increase in signal, when including the redshift distortions, for galaxy population at redshift $z=0.6$ (dashed line) and at redshift $z=1.7$ (solid line).  The relative increase is larger (over 40\% compared to over 30\%) at high redshifts for two reasons: the distortion parameter which modulates the amplitude of the redshift distortion is larger at high redshift, and the undistorted ISW signal is smaller. The effect of redshift distortions on parameter estimation is shown in Table \ref{redshift:tab}.}
\label{redshift:fig:comparison}
\end{figure}

In fact this new dependence can help probe alternative models of gravity as well as anisotropic stress.  The first term $\left(\frac{\rd \ln D(z)}{\rd \ln a}-1\right)$ is due to the change in gravitational potential and is related to the sum of both Newtonian potentials $\psi$ and $\Phi$, whereas the redshift distortion term, $\beta$, is a tracer of only the temporal potential $\psi$.  This means that if cosmic variance were not so large, the ISW signal including redshift distortions could potentially probe both potentials on its own.
\begin{table}
\caption{Constraints on $\Omega_\Lambda$ using different theoretical models to predict the ISW signal.  The original input is a model with $\Omega_\Lambda^0 = 0.70, \Omega_k=0$ and includes the effect of redshift distortions.  Omitting the effect of redshift distortions in the theoretical calculations mean we overestimate the true value of $\Omega_\Lambda^0$ while understimating error bars.   Using the Limber equations means one will underestimate the value of $\Omega_\Lambda^0$.  All fits are consistent with each other at the 1$\sigma$ level.  Model 1 is calculated using Eq. \ref{limber:eq:cgt}. Model 2 using Eq. \ref{limber:eq:cgtzz}.  Model 3 with Eq. \ref{linear:eq:cgt} and \ref{linear:eq:wgreal}.  Model 4 using Eq. \ref{linear:eq:cgt} and \ref{redshift:eq:wgz:pad}.}
\begin{center}
\begin{tabular}{cccc}
\hline
&Theoretical Model& $z=0.1$&$z=0.6$\\
\hline  \\

1&Small Angle & $0.67^{+0.16}_{-0.24}$&$0.64^{+0.15}_{-0.33}$\\
&No Redshift Distortions&\\
&&\\
2&Small Angle & $0.67^{+0.17}_{-0.27}$&$0.61^{+0.17}_{-0.30}$\\
&With Redshift Distortions &\\
&\\
3&Exact Without&$0.73^{+0.15}_{-0.26}$&$0.73^{+0.12}_{-0.31}$\\
&Redshift Distortions&\\
&\\
4&Exact With&$0.70^{+0.20}_{-0.36}$&$0.70^{+0.19}_{-0.39}$\\
&Redshift Distortions&\\

\hline
\\
Original & $\Omega_\Lambda^0 = 0.70$\\
Input &\\
\hline
\end{tabular}
\end{center}
\label{redshift:tab}
\end{table}%
In Figure \ref{redshift:fig:comparison} we compare the ISW signal with and without the effect of redshift distortions.  The effect of the redshift distortions is to increase the ISW signal at all multipoles where the ISW signal is not negligible.  In the top panel of Figure \ref{redshift:fig:comparison} we show the signal at two different redshifts.  In the bottom panel of Figure \ref{redshift:fig:comparison} we show the relative increase in signal due to redshift distortions.  This relative increase can be over 30\% on the largest scales for a cross-correlation with a galaxy sample at redshift $z=0.6$.  For a cross-correlation with a galaxy sample at redshift $z=1.7$, the relative increase in signal is even larger (over 40\% at the largest scales), because the distortion parameter $\beta$ which modulates the amplitude of the redshift distortion effect is larger at high redshift, and the undistorted ISW signal is smaller.  At lower redshifts, the effect of redshift distortions is smaller yet non-negligible ($<10\%$ at $z=0.1$).


As the ISW cross-correlation signal will be affected by redshift distortions, so will the corresponding error bars.  The covariance of the cross-correlation ISW signal is given by: 
\be \Delta^2 C_{\ell,gT}=\frac{1}{(2\ell+1)f_{\rm sky}}\left[ C_{\ell, gT}^2+C_{\ell,gg}C_{\ell,TT}\right].\ee  
In the presence of redshift distortions, both terms $C_{\ell, gT}$ and $C_{\ell,gg}$ should be modified to include the redshift space galaxy window function.  The effect of the redshift distortions is to increase power in both terms.

We find that the signal to noise (i.e., $C_{gT}/\Delta C_{gT}$) is decreased by about $10\%$ ($15\%$) at redshifts $z=0.6$ ($1.7$), when redshift distortions are included.  However, the extra feature introduced by redshift distortions will help break cosmological parameter degeneracies, especially those involving $\Omega_m, w_0, b(z)$ and $\Omega_\Lambda$ (in a flat universe). 

In order to investigate the effect of using the undistorted ISW equation for a measurement that includes redshift distortions, we fit a mock ISW signal (including redshift distortions) using different theoretical equations.  The results for this exercise are given in Table \ref{redshift:tab}. We find that omitting the effect of the redshift distortions means we overestimate the true value of $\Omega_\Lambda^0$, and underestimate the size of the statistical error bars.  
This may explain in part fits to ISW spectra in the literature which prefer high dark energy density values, though only if the selection functions used were estimated from the galaxy redshift distribution.  For such surveys,  the results from Table \ref{redshift:tab} show the effect of redshift distortions should be included when constraining cosmological parameters from a galaxy-temperature cross-correlation measurement. 

\section{ISW Limber Equations with Redshift Distortions}\label{limber}
The ISW effect is a large scale effect, but for high redshift surveys and for larger multipoles 
it is often assumed that it is acceptable to use the small angle approximation or \emph{Limber equation} \citep{Rassat:2007krl,Afshordi:2003xu,Ho:2008}.  This uses the fact that on small angles (or large $\ell$): \be \lim_{\ell \to \infty}j_\ell(kr) =
\sqrt{\frac{\pi}{2\ell+1}}\delta\left(\ell+1/2 - kr\right)\label{limber:eq:jl}\ee
In the real space limit (i.e. using Eq. \ref{linear:eq:wgreal} as the galaxy window function), the undistorted ISW signal can be written in the small angle approximation by:  \be C_{\rg T}(\ell) = A_k\int \rd r
\varphi(r)D^2H\left[\frac{\rm{dln}D}{\rm{dln}a}-1\right]P\left(\frac{\ell+1/2}{r}\right). \label{limber:eq:cgt}
\ee

We find here a similar \emph{Limber equation} for the ISW signal including the redshift distortions.  It is given by: 
\bea 
\lefteqn{C_{\rg T}(\ell) =A_k \times}\label{limber:eq:cgtzz}\\
& \int \rd r
\varphi(r)D^2H\left[\frac{\rm{dln}D}{\rm{dln}a}-1\right]\left[1+\beta\left(A_\ell-B_\ell-D_\ell\right)\right]P\left(\frac{\ell+1/2}{r}\right) \nonumber 
,\eea which makes the extended assumption that $\lim_{\ell \to \infty}W_{g,\ell-2}(k) = \lim_{\ell \to \infty}W_{g,\ell+2}(k) = W_{g,\ell}(k).$

\begin{figure}
\includegraphics[width=8cm]{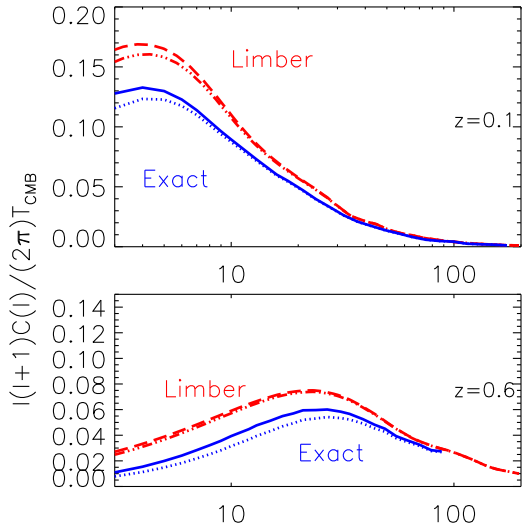}
\caption{Comparison of different equation for the ISW cross-power signal.  In blue are plotted the exact equations with (solid) and without (dotted) redshift distortions. In red are plotted the Limber equations with (dashed) and without (dot-dot-dashed) redshift distortions.  These are plotted for two galaxy populations with bins centered at redshifts $z=0.6$ and $z=1.7$.  At both redshifts the Limber equation overestimate the true value of the ISW signal.  The effect of this on parameter estimation is shown in Table \ref{redshift:tab}.}
\label{limber:fig:comparison}
\end{figure}

In Figure \ref{limber:fig:comparison} we compare the ISW cross-correlation signal, with and without redshift distortions, for the exact prescription and for the Limber approximations.  The galaxy distribution is centred at redshift $z=0.1$ and $z=0.6$ with width $\Delta z = 0.1$.  We find that  for the undistorted and redshift distorted cases, the Limber approximation overstimates the ISW signal by up to 30\% at $z=0.1$ and 80\% at $z=0.6$ for the widths considered.  This discrepency is largely due to the relatively small redshift bin width.  We find that for larger bin widths, the discrepency between the Limber and the exact equations decreases, as suggested by \cite{Loverde:2008limber}.  We also find that for a fixed bin width, the difference between the exact and Limber equations at a given $\ell$ increases with redshift, since the Limber approximation assumes $\ell +1/2 \sim kr$ (see Eq. \ref{limber:eq:jl}).

In Table \ref{redshift:tab} we show the effect of using the ISW Limber equations to constrain the dark energy density; we find that at all redshifts the result is to understimate the dark energy density, with the discrepency being larger at high redshifts.  For wide redshift bins this effect should be smaller.

\section{Conclusion}\label{ccl}
The ISW effect is weak compared to the signal from galaxy clustering or weak lensing, yet its simplicity as a linear effect and its potential with future tomographic surveys make it a promising tool to study.  As a direct tracer of the evolution of large scale gravitational potentials it is ideal for studying dark energy, curvature and departures from general relativity on large scales.

In this paper we show that the ISW signal is not as featureless as previously thought.  We show that the redshift distortions due to galaxy peculiar velocities affect the \emph{galaxy}-Temperature ISW signal, when the galaxy selection function is estimated from the redshift distribution.  As far as we are aware this is the first time that this effect has been included in the ISW signal.

We find that the effect of redshift distortions is to add power to the ISW signal.  At redshift $z=0.6$ this increase can be over 30\% on large scales.  The effect becomes larger at higher redshifts where the redshift distortion parameter $\beta$ is large and the amplitude of the ISW signal is small.  As well as adding power on large scales, the redshift distortions will also increase the covariance on the ISW cross-correlation power, though the additional feature should help break degeneracies between cosmological parameters.


The combination of the redshift distortion effect in the ISW signal is also useful to probe alternative models of gravity as the galaxy-Temperature correlation now probes both the sum of the Newtonian potentials $\psi + \Phi$ as well as the temporal potential $\psi$ separately.

If an ISW measurement includes the effect of redshift distortions, omitting this in the analysis would lead to an overestimation of the dark energy density $\Omega_\Lambda$, as well as an underestimation of its error bars. This may explain in part fits to ISW spectra in the literature which prefer high dark energy density values, though only if the selection functions used were estimated from the galaxy redshift distribution. 

We also find a new ISW Limber equation which includes redshift distortions.  We find that Limber equations in general overestimate the ISW signal at all redshifts, and that the effect is large for thin tomographic bins and at high redshifts.  Use of the ISW Limber approximation is such cases to constrain cosmology will lead to an underestimation of the true value of $\Omega_\Lambda$. 

The code used in this paper is available on request and will soon be added as a module to the iCosmo platform (http://www.icosmo.org).





\section*{Acknowledgments} 
The author thanks Alexandre R\'efr\'egier, Marian Douspis, Nabila Aghanim, Ofer Lahav, Niayesh Afshordi, Olivier Dor\'e, Sarah Bridle, and Martin Kunz for useful discussions.

\bibliographystyle{mn2e}
\bibliography{references}

\begin{thebibliography}{30}
\expandafter\ifx\csname natexlab\endcsname\relax\def\natexlab#1{#1}\fi
\expandafter\ifx\csname bibnamefont\endcsname\relax
  \def\bibnamefont#1{#1}\fi
\expandafter\ifx\csname bibfnamefont\endcsname\relax
  \def\bibfnamefont#1{#1}\fi
\expandafter\ifx\csname citenamefont\endcsname\relax
  \def\citenamefont#1{#1}\fi
\expandafter\ifx\csname url\endcsname\relax
  \def\url#1{\texttt{#1}}\fi
\expandafter\ifx\csname urlprefix\endcsname\relax\def\urlprefix{URL }\fi
\providecommand{\bibinfo}[2]{#2}
\providecommand{\eprint}[2][]{\url{#2}}

\bibitem[{\citenamefont{Sachs and Wolfe}(1967)}]{Sachs:1967er}
\bibinfo{author}{\bibfnamefont{R.~K.} \bibnamefont{Sachs}} \bibnamefont{and}
  \bibinfo{author}{\bibfnamefont{A.~M.} \bibnamefont{Wolfe}},
  \bibinfo{journal}{Astrophys. J.} \textbf{\bibinfo{volume}{147}},
  \bibinfo{pages}{73} (\bibinfo{year}{1967}).

\bibitem[{\citenamefont{{Douspis} et~al.}(2008)\citenamefont{{Douspis},
  {Castro}, {Caprini}, and {Aghanim}}}]{douspis:2008}
\bibinfo{author}{\bibfnamefont{M.}~\bibnamefont{{Douspis}}},
  \bibinfo{author}{\bibfnamefont{P.~G.} \bibnamefont{{Castro}}},
  \bibinfo{author}{\bibfnamefont{C.}~\bibnamefont{{Caprini}}},
  \bibnamefont{and}
  \bibinfo{author}{\bibfnamefont{N.}~\bibnamefont{{Aghanim}}},
  \bibinfo{journal}{AAP} \textbf{\bibinfo{volume}{485}}, \bibinfo{pages}{395}
  (\bibinfo{year}{2008}), \eprint{0802.0983}.

\bibitem[{\citenamefont{{R{\'e}fr{\'e}gier} and {the DUNE
  collaboration}}(2008)}]{refregier:2008}
\bibinfo{author}{\bibfnamefont{A.}~\bibnamefont{{R{\'e}fr{\'e}gier}}}
  \bibnamefont{and} \bibinfo{author}{\bibnamefont{{the DUNE collaboration}}},
  \bibinfo{journal}{ArXiv e-prints} \textbf{\bibinfo{volume}{802}}
  (\bibinfo{year}{2008}), \eprint{0802.2522}.

\bibitem[{\citenamefont{Fosalba et~al.}(2003)\citenamefont{Fosalba,
  Gazta\~naga, and Castander}}]{Fosalba:2003ge}
\bibinfo{author}{\bibfnamefont{P.}~\bibnamefont{Fosalba}},
  \bibinfo{author}{\bibfnamefont{E.}~\bibnamefont{Gazta\~naga}},
  \bibnamefont{and}
  \bibinfo{author}{\bibfnamefont{F.}~\bibnamefont{Castander}},
  \bibinfo{journal}{Astrophys. J.} \textbf{\bibinfo{volume}{597}},
  \bibinfo{pages}{L89} (\bibinfo{year}{2003}), \eprint{astro-ph/0307249}.

\bibitem[{\citenamefont{Boughn and Crittenden}(2005)}]{Boughn:2004zm}
\bibinfo{author}{\bibfnamefont{S.~P.} \bibnamefont{Boughn}} \bibnamefont{and}
  \bibinfo{author}{\bibfnamefont{R.~G.} \bibnamefont{Crittenden}},
  \bibinfo{journal}{New Astron. Rev.} \textbf{\bibinfo{volume}{49}},
  \bibinfo{pages}{75} (\bibinfo{year}{2005}), \eprint{astro-ph/0404470}.

\bibitem[{\citenamefont{Afshordi et~al.}(2004)\citenamefont{Afshordi, Loh, and
  Strauss}}]{Afshordi:2003xu}
\bibinfo{author}{\bibfnamefont{N.}~\bibnamefont{Afshordi}},
  \bibinfo{author}{\bibfnamefont{Y.-S.} \bibnamefont{Loh}}, \bibnamefont{and}
  \bibinfo{author}{\bibfnamefont{M.~A.} \bibnamefont{Strauss}},
  \bibinfo{journal}{Phys. Rev.} \textbf{\bibinfo{volume}{D69}},
  \bibinfo{pages}{083524} (\bibinfo{year}{2004}), \eprint{astro-ph/0308260}.

\bibitem[{\citenamefont{Nolta et~al.}(2004)}]{Nolta:2003uy}
\bibinfo{author}{\bibfnamefont{M.~R.} \bibnamefont{Nolta}}
  \bibnamefont{et~al.}, \bibinfo{journal}{Astrophys. J.}
  \textbf{\bibinfo{volume}{608}}, \bibinfo{pages}{10} (\bibinfo{year}{2004}),
  \eprint{astro-ph/0305097}.

\bibitem[{\citenamefont{Padmanabhan et~al.}(2005)}]{Padmanabhan:2004fy}
\bibinfo{author}{\bibfnamefont{N.}~\bibnamefont{Padmanabhan}}
  \bibnamefont{et~al.}, \bibinfo{journal}{Phys. Rev.}
  \textbf{\bibinfo{volume}{D72}}, \bibinfo{pages}{043525}
  (\bibinfo{year}{2005}), \eprint{astro-ph/0410360}.

\bibitem[{\citenamefont{{Cabr{\'e}} et~al.}(2006)\citenamefont{{Cabr{\'e}},
  {Gazta{\~n}aga}, {Manera}, {Fosalba}, and {Castander}}}]{Cabre:2006qm}
\bibinfo{author}{\bibfnamefont{A.}~\bibnamefont{{Cabr{\'e}}}},
  \bibinfo{author}{\bibfnamefont{E.}~\bibnamefont{{Gazta{\~n}aga}}},
  \bibinfo{author}{\bibfnamefont{M.}~\bibnamefont{{Manera}}},
  \bibinfo{author}{\bibfnamefont{P.}~\bibnamefont{{Fosalba}}},
  \bibnamefont{and}
  \bibinfo{author}{\bibfnamefont{F.}~\bibnamefont{{Castander}}},
  \bibinfo{journal}{MNRAS} \textbf{\bibinfo{volume}{372}}, \bibinfo{pages}{L23}
  (\bibinfo{year}{2006}), \eprint{arXiv:astro-ph/0603690}.

\bibitem[{\citenamefont{Gazta\~naga et~al.}(2006)\citenamefont{Gazta\~naga,
  Manera, and Multamaki}}]{Gaztanaga:2004sk}
\bibinfo{author}{\bibfnamefont{E.}~\bibnamefont{Gazta\~naga}},
  \bibinfo{author}{\bibfnamefont{M.}~\bibnamefont{Manera}}, \bibnamefont{and}
  \bibinfo{author}{\bibfnamefont{T.}~\bibnamefont{Multamaki}},
  \bibinfo{journal}{Mon. Not. Roy. Astron. Soc.}
  \textbf{\bibinfo{volume}{365}}, \bibinfo{pages}{171} (\bibinfo{year}{2006}),
  \eprint{astro-ph/0407022}.

\bibitem[{\citenamefont{{Giannantonio}
  et~al.}(2006)\citenamefont{{Giannantonio}, {Crittenden}, {Nichol},
  {Scranton}, {Richards}, {Myers}, {Brunner}, {Gray}, {Connolly}, and
  {Schneider}}}]{Giannantonio:2006al}
\bibinfo{author}{\bibfnamefont{T.}~\bibnamefont{{Giannantonio}}},
  \bibinfo{author}{\bibfnamefont{R.~G.} \bibnamefont{{Crittenden}}},
  \bibinfo{author}{\bibfnamefont{R.~C.} \bibnamefont{{Nichol}}},
  \bibinfo{author}{\bibfnamefont{R.}~\bibnamefont{{Scranton}}},
  \bibinfo{author}{\bibfnamefont{G.~T.} \bibnamefont{{Richards}}},
  \bibinfo{author}{\bibfnamefont{A.~D.} \bibnamefont{{Myers}}},
  \bibinfo{author}{\bibfnamefont{R.~J.} \bibnamefont{{Brunner}}},
  \bibinfo{author}{\bibfnamefont{A.~G.} \bibnamefont{{Gray}}},
  \bibinfo{author}{\bibfnamefont{A.~J.} \bibnamefont{{Connolly}}},
  \bibnamefont{and} \bibinfo{author}{\bibfnamefont{D.~P.}
  \bibnamefont{{Schneider}}}, \bibinfo{journal}{PRD}
  \textbf{\bibinfo{volume}{74}}, \bibinfo{pages}{063520}
  (\bibinfo{year}{2006}), \eprint{arXiv:astro-ph/0607572}.

\bibitem[{\citenamefont{{McEwen} et~al.}(2007)\citenamefont{{McEwen}, {Vielva},
  {Hobson}, {Mart{\'{\i}}nez-Gonz{\'a}lez}, and {Lasenby}}}]{McEwen:2006md}
\bibinfo{author}{\bibfnamefont{J.~D.} \bibnamefont{{McEwen}}},
  \bibinfo{author}{\bibfnamefont{P.}~\bibnamefont{{Vielva}}},
  \bibinfo{author}{\bibfnamefont{M.~P.} \bibnamefont{{Hobson}}},
  \bibinfo{author}{\bibfnamefont{E.}~\bibnamefont{{Mart{\'{\i}}nez-Gonz{\'a}le%
z}}}, \bibnamefont{and} \bibinfo{author}{\bibfnamefont{A.~N.}
  \bibnamefont{{Lasenby}}}, \bibinfo{journal}{MNRAS}
  \textbf{\bibinfo{volume}{376}}, \bibinfo{pages}{1211} (\bibinfo{year}{2007}),
  \eprint{arXiv:astro-ph/0602398}.

\bibitem[{\citenamefont{{Rassat} et~al.}(2007)\citenamefont{{Rassat}, {Land},
  {Lahav}, and {Abdalla}}}]{Rassat:2007krl}
\bibinfo{author}{\bibfnamefont{A.}~\bibnamefont{{Rassat}}},
  \bibinfo{author}{\bibfnamefont{K.}~\bibnamefont{{Land}}},
  \bibinfo{author}{\bibfnamefont{O.}~\bibnamefont{{Lahav}}}, \bibnamefont{and}
  \bibinfo{author}{\bibfnamefont{F.~B.} \bibnamefont{{Abdalla}}},
  \bibinfo{journal}{MNRAS} \textbf{\bibinfo{volume}{377}},
  \bibinfo{pages}{1085} (\bibinfo{year}{2007}).

\bibitem[{\citenamefont{{Giannantonio}
  et~al.}(2008)\citenamefont{{Giannantonio}, {Scranton}, {Crittenden},
  {Nichol}, {Boughn}, {Myers}, and {Richards}}}]{Giannantonio:2008}
\bibinfo{author}{\bibfnamefont{T.}~\bibnamefont{{Giannantonio}}},
  \bibinfo{author}{\bibfnamefont{R.}~\bibnamefont{{Scranton}}},
  \bibinfo{author}{\bibfnamefont{R.~G.} \bibnamefont{{Crittenden}}},
  \bibinfo{author}{\bibfnamefont{R.~C.} \bibnamefont{{Nichol}}},
  \bibinfo{author}{\bibfnamefont{S.~P.} \bibnamefont{{Boughn}}},
  \bibinfo{author}{\bibfnamefont{A.~D.} \bibnamefont{{Myers}}},
  \bibnamefont{and} \bibinfo{author}{\bibfnamefont{G.~T.}
  \bibnamefont{{Richards}}}, \bibinfo{journal}{PRD}
  \textbf{\bibinfo{volume}{77}}, \bibinfo{pages}{123520}
  (\bibinfo{year}{2008}), \eprint{0801.4380}.

\bibitem[{\citenamefont{Crittenden and Turok}(1996)}]{Crittenden:1995ak}
\bibinfo{author}{\bibfnamefont{R.~G.} \bibnamefont{Crittenden}}
  \bibnamefont{and} \bibinfo{author}{\bibfnamefont{N.}~\bibnamefont{Turok}},
  \bibinfo{journal}{Phys. Rev. Lett.} \textbf{\bibinfo{volume}{76}},
  \bibinfo{pages}{575} (\bibinfo{year}{1996}), \eprint{astro-ph/9510072}.

\bibitem[{\citenamefont{{Loverde} et~al.}(2007)\citenamefont{{Loverde}, {Hui},
  and {Gazta{\~n}aga}}}]{Loverde:2007}
\bibinfo{author}{\bibfnamefont{M.}~\bibnamefont{{Loverde}}},
  \bibinfo{author}{\bibfnamefont{L.}~\bibnamefont{{Hui}}}, \bibnamefont{and}
  \bibinfo{author}{\bibfnamefont{E.}~\bibnamefont{{Gazta{\~n}aga}}},
  \bibinfo{journal}{prd} \textbf{\bibinfo{volume}{75}}, \bibinfo{pages}{043519}
  (\bibinfo{year}{2007}).

\bibitem[{\citenamefont{{Fisher} et~al.}(1994)\citenamefont{{Fisher}, {Scharf},
  and {Lahav}}}]{Fisher:1994}
\bibinfo{author}{\bibfnamefont{K.~B.} \bibnamefont{{Fisher}}},
  \bibinfo{author}{\bibfnamefont{C.~A.} \bibnamefont{{Scharf}}},
  \bibnamefont{and} \bibinfo{author}{\bibfnamefont{O.}~\bibnamefont{{Lahav}}},
  \bibinfo{journal}{MNRAS} \textbf{\bibinfo{volume}{266}}, \bibinfo{pages}{219}
  (\bibinfo{year}{1994}), \eprint{astro-ph/9309027}.

\bibitem[{\citenamefont{{Heavens} and {Taylor}}(1995)}]{Heavens:1995}
\bibinfo{author}{\bibfnamefont{A.~F.} \bibnamefont{{Heavens}}}
  \bibnamefont{and} \bibinfo{author}{\bibfnamefont{A.~N.}
  \bibnamefont{{Taylor}}}, \bibinfo{journal}{MNRAS}
  \textbf{\bibinfo{volume}{275}}, \bibinfo{pages}{483} (\bibinfo{year}{1995}),
  \eprint{arXiv:astro-ph/9409027}.

\bibitem[{\citenamefont{{Guzzo} and et~al.}(2008)}]{Guzzo:2008}
\bibinfo{author}{\bibfnamefont{L.}~\bibnamefont{{Guzzo}}} \bibnamefont{and}
  \bibinfo{author}{\bibnamefont{et~al.}}, \bibinfo{journal}{Nature}
  \textbf{\bibinfo{volume}{451}}, \bibinfo{pages}{541} (\bibinfo{year}{2008}),
  \eprint{0802.1944}.

\bibitem[{\citenamefont{{Fosalba} and {Dor{\'e}}}(2007)}]{Dore:2007}
\bibinfo{author}{\bibfnamefont{P.}~\bibnamefont{{Fosalba}}} \bibnamefont{and}
  \bibinfo{author}{\bibfnamefont{O.}~\bibnamefont{{Dor{\'e}}}},
  \bibinfo{journal}{PRD} \textbf{\bibinfo{volume}{76}}, \bibinfo{pages}{103523}
  (\bibinfo{year}{2007}), \eprint{arXiv:astro-ph/0701782}.

\bibitem[{\citenamefont{{Barreiro} et~al.}(2008)\citenamefont{{Barreiro},
  {Vielva}, {Hernandez-Monteagudo}, and {Martinez-Gonzalez}}}]{iswmap:2008}
\bibinfo{author}{\bibfnamefont{R.~B.} \bibnamefont{{Barreiro}}},
  \bibinfo{author}{\bibfnamefont{P.}~\bibnamefont{{Vielva}}},
  \bibinfo{author}{\bibfnamefont{C.}~\bibnamefont{{Hernandez-Monteagudo}}},
  \bibnamefont{and}
  \bibinfo{author}{\bibfnamefont{E.}~\bibnamefont{{Martinez-Gonzalez}}},
  \bibinfo{journal}{IEEE Journal of Selected Topics in Signal Processing, vol
  2, issue 5, p.~747-754} \textbf{\bibinfo{volume}{5}}, \bibinfo{pages}{747}
  (\bibinfo{year}{2008}), \eprint{0809.2557}.

\bibitem[{\citenamefont{{Granett} et~al.}(2008)\citenamefont{{Granett},
  {Neyrinck}, and {Szapudi}}}]{granett:2008}
\bibinfo{author}{\bibfnamefont{B.~R.} \bibnamefont{{Granett}}},
  \bibinfo{author}{\bibfnamefont{M.~C.} \bibnamefont{{Neyrinck}}},
  \bibnamefont{and}
  \bibinfo{author}{\bibfnamefont{I.}~\bibnamefont{{Szapudi}}},
  \bibinfo{journal}{ArXiv e-prints}  (\bibinfo{year}{2008}),
  \eprint{0812.1025}.

\bibitem[{\citenamefont{{Kamionkowski} and
  {Spergel}}(1994)}]{Kamionkowski:1994s}
\bibinfo{author}{\bibfnamefont{M.}~\bibnamefont{{Kamionkowski}}}
  \bibnamefont{and} \bibinfo{author}{\bibfnamefont{D.~N.}
  \bibnamefont{{Spergel}}}, \bibinfo{journal}{APJ}
  \textbf{\bibinfo{volume}{432}}, \bibinfo{pages}{7} (\bibinfo{year}{1994}),
  \eprint{astro-ph/9312017}.

\bibitem[{\citenamefont{{Kinkhabwala} and
  {Kamionkowski}}(1999)}]{Kinkhabwala:1999k}
\bibinfo{author}{\bibfnamefont{A.}~\bibnamefont{{Kinkhabwala}}}
  \bibnamefont{and}
  \bibinfo{author}{\bibfnamefont{M.}~\bibnamefont{{Kamionkowski}}},
  \bibinfo{journal}{Physical Review Letters} \textbf{\bibinfo{volume}{82}},
  \bibinfo{pages}{4172} (\bibinfo{year}{1999}), \eprint{astro-ph/9808320}.

\bibitem[{\citenamefont{{Kaiser}}(1987)}]{Kaiser:1987}
\bibinfo{author}{\bibfnamefont{N.}~\bibnamefont{{Kaiser}}},
  \bibinfo{journal}{MNRAS} \textbf{\bibinfo{volume}{227}}, \bibinfo{pages}{1}
  (\bibinfo{year}{1987}).

\bibitem[{\citenamefont{{Padmanabhan} and al.}(2007)}]{Padmanabhan:2006}
\bibinfo{author}{\bibfnamefont{N.~e.} \bibnamefont{{Padmanabhan}}}
  \bibnamefont{and} \bibinfo{author}{\bibnamefont{al.}},
  \bibinfo{journal}{MNRAS} \textbf{\bibinfo{volume}{378}}, \bibinfo{pages}{852}
  (\bibinfo{year}{2007}), \eprint{arXiv:astro-ph/0605302}.

\bibitem[{\citenamefont{{Rassat} et~al.}(2008)\citenamefont{{Rassat}, {Amara},
  {Amendola}, {Castander}, {Kitching}, {Kunz}, {Refregier}, {Wang}, and
  {Weller}}}]{Rassat:2008bao}
\bibinfo{author}{\bibfnamefont{A.}~\bibnamefont{{Rassat}}},
  \bibinfo{author}{\bibfnamefont{A.}~\bibnamefont{{Amara}}},
  \bibinfo{author}{\bibfnamefont{L.}~\bibnamefont{{Amendola}}},
  \bibinfo{author}{\bibfnamefont{F.~J.} \bibnamefont{{Castander}}},
  \bibinfo{author}{\bibfnamefont{T.}~\bibnamefont{{Kitching}}},
  \bibinfo{author}{\bibfnamefont{M.}~\bibnamefont{{Kunz}}},
  \bibinfo{author}{\bibfnamefont{A.}~\bibnamefont{{Refregier}}},
  \bibinfo{author}{\bibfnamefont{Y.}~\bibnamefont{{Wang}}}, \bibnamefont{and}
  \bibinfo{author}{\bibfnamefont{J.}~\bibnamefont{{Weller}}},
  \bibinfo{journal}{ArXiv e-prints}  (\bibinfo{year}{2008}),
  \eprint{0810.0003}.

\bibitem[{\citenamefont{{Peebles} and {Yu}}(1970)}]{Peebles:1970Y}
\bibinfo{author}{\bibfnamefont{P.~J.~E.} \bibnamefont{{Peebles}}}
  \bibnamefont{and} \bibinfo{author}{\bibfnamefont{J.~T.} \bibnamefont{{Yu}}},
  \bibinfo{journal}{Apj} \textbf{\bibinfo{volume}{162}}, \bibinfo{pages}{815}
  (\bibinfo{year}{1970}).

\bibitem[{\citenamefont{{Ho} et~al.}(2008)\citenamefont{{Ho}, {Hirata},
  {Padmanabhan}, {Seljak}, and {Bahcall}}}]{Ho:2008}
\bibinfo{author}{\bibfnamefont{S.}~\bibnamefont{{Ho}}},
  \bibinfo{author}{\bibfnamefont{C.}~\bibnamefont{{Hirata}}},
  \bibinfo{author}{\bibfnamefont{N.}~\bibnamefont{{Padmanabhan}}},
  \bibinfo{author}{\bibfnamefont{U.}~\bibnamefont{{Seljak}}}, \bibnamefont{and}
  \bibinfo{author}{\bibfnamefont{N.}~\bibnamefont{{Bahcall}}},
  \bibinfo{journal}{\prd} \textbf{\bibinfo{volume}{78}},
  \bibinfo{pages}{043519} (\bibinfo{year}{2008}), \eprint{0801.0642}.

\bibitem[{\citenamefont{{Loverde} and {Afshordi}}(2008)}]{Loverde:2008limber}
\bibinfo{author}{\bibfnamefont{M.}~\bibnamefont{{Loverde}}} \bibnamefont{and}
  \bibinfo{author}{\bibfnamefont{N.}~\bibnamefont{{Afshordi}}},
  \bibinfo{journal}{PRD} \textbf{\bibinfo{volume}{78}}, \bibinfo{pages}{123506}
  (\bibinfo{year}{2008}), \eprint{0809.5112}.

\end{thebibliography}



\end{document}